\pdfoutput=1

\documentclass[11pt]{article}

\usepackage[]{acl}
\usepackage{placeins}
\usepackage{float}
\usepackage{booktabs} 

\usepackage{times}
\usepackage{latexsym}

\usepackage[T1]{fontenc}

\usepackage[utf8]{inputenc}

\usepackage{microtype}
\usepackage{graphicx}
\usepackage{verbatim}
\usepackage{tikz}
\usetikzlibrary{positioning, shapes.geometric, arrows.meta}
\usepackage{listings}
\usepackage{xcolor}
\lstdefinestyle{pythonstyle}{
language=Python,
backgroundcolor=\color[rgb]{0.97,0.97,0.97},
basicstyle=\ttfamily\tiny,
keywordstyle=\color[rgb]{0.0,0.45,0.68},
commentstyle=\color[rgb]{0.13,0.55,0.14},
stringstyle=\color[rgb]{0.66,0.05,0.57},
showspaces=false,
showstringspaces=false,
showtabs=true,
tabsize=2,
breaklines=true,
frame=single,
xleftmargin=2em,
framexleftmargin=2em,
aboveskip=1.5em,
belowskip=1.5em,
numbers=left,
numbersep=5pt,
numberstyle=\footnotesize\ttfamily\tiny\color[rgb]{0.4,0.4,0.4},
captionpos=b, 
upquote=true 
}
\title{Can Github issues be solved with Tree Of Thoughts?}

\author{Ricardo La Rosa \\
  \ttfamily\small\texttt{ricardo@larosa.dev} \\\And
  Corey Hulse \\
  \ttfamily\small\texttt{coreyohulse@gmail.com} \\\And
  Bangdi Liu \\ 
  \ttfamily\small\texttt{buddytt0915@gmail.com}
}

\begin{document}
\maketitle
\begin{abstract}
While there have been extensive studies in code generation by large language models (LLM), where benchmarks like HumanEval\cite{chen2021codex} have been surpassed with an impressive 96.3\% success rate, these benchmarks predominantly judge a model's performance on basic function-level code generation and lack the critical thinking and concept of scope required of real-world scenarios such as solving GitHub issues. This research introduces the application of the Tree of Thoughts (ToT) \cite{yao2023tree} language model reasoning framework for enhancing the decision-making and problem-solving abilities of LLMs for this complex task. Compared to traditional input-output (IO) prompting and Retrieval Augmented Generation (RAG) techniques, ToT is designed to improve performance by facilitating a structured exploration of multiple reasoning trajectories and enabling self-assessment of potential solutions. We experimentally deploy ToT in tackling a Github issue contained within an \textit{instance} of the SWE-bench\cite{jimenez2024swebench}. \\
However, our results reveal that the ToT framework alone is not enough to give LLMs the critical reasoning capabilities to outperform existing methods. In this paper we analyze the potential causes of these shortcomings and identify key areas for improvement such as deepening the thought process and introducing \textit{agentic}\cite{Ng2024} capabilities. The insights of this research are aimed at informing future directions for refining the application of ToT and better harnessing the potential of LLMs in real-world problem-solving scenarios.
\end{abstract}

\section{Introduction}
Tree of Thoughts (ToT) \cite{yao2023tree} is a language model reasoning framework designed to enhance the autonomy and intelligence of language models (LMs) in decision-making and problem-solving tasks. ToT managed to outperform Input-Output prompting (IO), Chain of Thought (CoT) \cite{Wei2022ChainOfThought}, and Self Consistency with CoT (CoT-SC) in several reasoning based tasks such as \textit{Game of 24}, \textit{Crosswords} and \textit{Creative Writing}. Despite the promising results of ToT on these basic tasks, there is an absence of studies that apply ToT towards more complex tasks that more closely model the real-world. Our research aims to put this framework to test in one of the most challenging software engineering tasks for large language models: resolving GitHub issues. This task requires an overall sense of scope and understanding of the repository when making changes, which requires stronger critical reasoning skills than previous basic code generation tasks. We anticipate that the ToT method will outperform both IO prompting and Retrieval Augmented Generation (RAG) techniques in this task. This expectation is based on ToT's ability to instill a stronger ability in LLMS for decision-making, evaluating multiple reasoning paths and self-assessing choices to determine the subsequent course of action.
\section{Related Work}
Prior work on solving Github issues has been done by the Princeton NLP team as part of SWE-Bench\cite{jimenez2024swebench}. They released two fine-tuned models, SWE-Llama 7B and SWE-Llama 13B based on CodeLlama\cite{roziereCodeLlamaOpen2023} with Retrieval Augmented Generation (RAG). Further work then was done with the introduction of SWE-agent\cite{yang2024sweagent} which is a large language model-based agent system that operates within an Agent-Computer Interface (ACI). Another recent work is LLM-Based Multi-Agent Framework for GitHub Issue ReSolution (MAGIS) \cite{Tao2024MAGIS} which introduces Multi-Agency whereby leveraging the collaboration of various agents with distinct roles in the planning and coding process to resolve GitHub issues. 
\section{Data}
\subsection{SWE-bench}
We utilized the dataset provided by SWE-Bench\cite{jimenez2024swebench} as the basis for the experiments.
SWE-bench is a benchmark for evaluating large language models on real world software issues collected from GitHub. Given a code-base and an issue, a language model is tasked with generating a code patch that resolves the described problem.
With SWE-Bench, you can:
\begin{itemize}
    \item Train or fine-tune a model with their pre-processed datasets.
    \item Run inference on existing models.
    \item Evaluate a model against the benchmark and determine the correctness of a solution proposed by the model.
\end{itemize}
This dataset is composed of a wide variety of tasks, such as filing a bug report or making a feature request, that the model will be charged with completing. The main similarity between these tasks is that they all require the model to generate a git patch to an existing code-base based on the problem statement of the Github issue. The revised code-base is then evaluated using the internal testing framework of the repository. If the proposed patch passes these tests then the model's proposed changes are considered successful and the task is counted as passed. 
\subsection{SWE-bench Lite}
In order to reduce costs we used the SWE-bench\_Lite\cite{jimenez2024swe_bench_lite} dataset which is a canonical subset of SWE-bench that has been curated to make evaluation less costly.
Instances from the original dataset that match the following criteria are not considered:
\begin{enumerate}
    \item Include images, external links, references to specific commits, and references to other pull requests
    \item Contain problem statements with fewer than 40 words
    \item Edit more than one file
    \item Have a gold patch with more than three edit hunks
    \item Create or delete files
    \item Contain tests with error messages checks
\end{enumerate}
After filtering out the instances who violated the above standards, the result is a smaller dataset of 23 instances in the \textit{dev} split and 300 instances in the \textit{test} split.
\subsection{Motivation}
Traditional benchmarks in Natural Language Processing (NLP) often focus on relatively short input and output sequences that are not representative of real-world tasks. 

\begin{table}[ht]
\centering
\caption{HumanEval Leaderboard}
\label{tab:humaneval_leaderboard}
\resizebox{\columnwidth}{!}{
    \begin{tabular}{lcc}
    \toprule
    \textbf{Model} & \textbf{Success Rate(\%)}  \\
    \midrule
    AgentCoder(GPT-4) \cite{huang2024agentcoder} & 96.3 \\
    LDB + Reflexion(GPT-3.5) \cite{zhong2024ldb} & 95.1 \\
    Language Agent Search Tree(GPT-4) \cite{zhou2023language} & 94.4 \\
    L2MAC(GPT-4) \cite{holt2024l2mac} & 90.2 \\
    \bottomrule
    \end{tabular}
}
\end{table}

As shown in table \ref{tab:humaneval_leaderboard} LLMs demonstrate remarkable performance on the HumanEval\cite{chen2021codex} benchmark.  However, this benchmark exhibits several notable weaknesses: scope limited to function-level code generation, lack of diversity by focusing mainly on algorithmic tasks, and a lack of contextual and environmental interaction. In contrast, we considered that SWE-bench, emphasizes tasks that adequately model real-world scenarios where the interdependencies of the code base as a whole must be take into account when generating new patches, and the testing framework is able to use a built in testing framework to evaluate if the model's code correctly fits into the existing code base.

\begin{table}[ht]
\centering
\caption{SWE-bench Lite Leaderboard}
\label{tab:swe_lite_leaderboard}
\resizebox{\columnwidth}{!}{
    \begin{tabular}{lcc}
    \toprule
    \textbf{Model} & \textbf{Success Rate(\%)}  \\
    \midrule
    SWE-agent + GPT 4 & 17.00 \\
    SWE-agent + Claude 3 Opus & 11.67 \\
    RAG + Claude 3 Opus & 4.00 \\
    RAG + GPT4 & 2.67 \\
    RAG + Claude 2 & 2.00 \\
    RAG + SWE-Llama 13B & 1.67 \\
    RAG + SWE-Llama 7B & 1.33 \\
    RAG + GPT 3.5 & 0.33 \\
    GPT 4 & 0.00 \\
    ChatGPT 3.5 & 0.00 \\
    \bottomrule
    \end{tabular}
}
\end{table}

Table \ref{tab:swe_lite_leaderboard} paints a different picture.
Even with Retrieval Augmented Generation, performance sees limited improvement due to the difficulties that LLMs face when handling long context inputs, notably in tasks like resolving GitHub issues at the repository level, where using large portions of the repository as input is impractical.

\section{Models}

We utilized three open-source models for this research. The first model was \textbf{CodeLlama 34B}. \cite{jimenez2024swebench} highlighted that variants of CodeLlama were not capable of following detailed instructions in order to make repository-wide code edits, and typically resorted to outputting placeholder responses or unrelated code. To address this issue, they performed supervised fine-tuning on the 7 billion-parameter and 13 billion-parameter variants. The resulting models were shown to be highly successful at maintaining specialized repositories and could be run on consumer hardware to resolve GitHub issues. Based on these observations, we opted for the 34 billion parameter version of CodeLlama, which had been quantized to 4-bit precision and fine-tuned using a select portion of the SWE-bench dataset.

\begin{table}[ht]
\centering
\caption{Models}

\begin{tabular}{lc}
\toprule
\textbf{Model} & \textbf{Patch generation} \\
& \textbf{strategy} \\
\midrule
CodeLlama 34B & Supervised fine-tuning \\
Mixtral-8x7B & In-context learning \\
Llama2 70B & In-context learning \\
Llama3 70B Instruct & In-context learning \\
\bottomrule
\end{tabular}
\label{tab:models}
\end{table}

Table \ref{tab:models} presents the models employed in this research and their respective patch generation strategies.
 We broadened our approach by adding three larger models: \textbf{Mixtral-8x7B} \cite{jiang2024mixtral}, which comprises eight 7-billion parameter models with a Sparsely-Gated Mixture-of-Experts layer (MoE) \cite{shazeer2017outrageously}, and the 70-billion parameter models \textbf{Llama2 70B} and \textbf{Llama3 70B Instruct}. \\
 Considering their in-context learning capabilities, we hypothesized that these larger models would correctly generate patches in the unified diff format when presented with few-shot examples.

\section{Methods}

\subsection{Baselines}
We use a standard input-output (IO) prompt with five-shot examples.
\subsection{Tree of Thoughts setup}

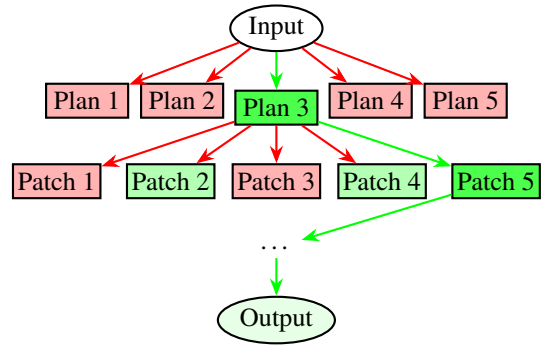
\begin{figure}[ht]
\centering
\begin{tikzpicture}[
every node/.style={scale=0.9}, 
base/.style={draw, thick, align=center, minimum width=1cm, minimum height=0.5cm, inner sep=2pt},
input/.style={base, ellipse},
thought/.style={base, rectangle},
output/.style={base, ellipse, fill=green!10},
best/.style={base, fill=green!70, minimum width=1.2cm},
bad/.style={base, fill=red!30, minimum width=1.2cm},
good/.style={base, fill=green!30, minimum width=1.2cm},
line/.style={-Stealth, thick},
greenline/.style={line, draw=green},
redline/.style={line, draw=red}
]
\node [input] (input) {Input};
\node [bad, below left=0.5cm and 1.5cm of input] (p1) {Plan 1};
\node [bad, below left=0.5cm and 0.25cm of input] (p3) {Plan 2};
\node [best, below=0.5cm of input] (x) {Plan 3};

\node [bad, below right=0.5cm and 0.25cm of input] (p4) {Plan 4};
\node [bad, below right=0.5cm and 1.5cm of input] (p5) {Plan 5};
\node [good, below left=0.5cm and 0.25cm of x] (c) {Patch 2};

\node [bad, below=0.5cm of x] (d) {Patch 3};
\node [bad, below left=0.5cm and 1.75cm of x] (e) {Patch 1};
\node [good, below right=0.5cm and 0.25cm of x] (f) {Patch 4};
\node [best, below right=0.5cm and 1.75cm of x] (g) {Patch 5};
\node [below=0.5cm of d] (dots) {\ldots};
\node [output, below=0.5cm of dots] (output) {Output};
\draw [greenline] (input) -- (x);
\draw [redline] (input) -- (p1);
\draw [redline] (input) -- (p3);
\draw [redline] (input) -- (p4);
\draw [redline] (input) -- (p5);

\draw [redline] (x) -- (c);
\draw [redline] (x) -- (d);
\draw [redline] (x) -- (e);
\draw [redline] (x) -- (f);
\draw [greenline] (x) -- (g);

\draw [greenline] (g) -- (dots);
\draw [greenline] (dots) -- (output);
\end{tikzpicture}
\caption{ToT setup with \textit{n = 5}, \textit{k = 5} and \textit{b = 1}}
\label{fig:tot}
\end{figure}

As shown in figure \ref{fig:tot} we built a ToT with depth \textit{d = 2} with one intermediate thought step. The input is a Chain of Thought style prompt where the model is asked to generate \textit{n} plans and votes for the best one, then similarly generate \textit{k} patches based on the best plan. In order to guarantee plan diversity we increased the temperature to a determined value \textit{t}. The last step is to rate the patches: the patch with the highest score is chosen. 
The breadth limit is always set to \textit{b = 1} as a consequence the breadth-first search (BFS) only maintains the most promising state per step in a greedy approach.\\

We forked the official Tree of Thoughts Github repository\cite{yao2023totllm} and created a branch to add a new \texttt{Task} class called \texttt{SWETask}. This class was designed to solve the \textit{instances} within the SWE-bench Lite dataset, based on the mentioned ToT setup.
\subsubsection{Prompting}
A zero-shot vote prompt was used to sample votes for plan selection and zero-shot score prompt is used to make patches scores. Example of prompts used are in appendix \ref{sec:appendix}.

\subsection{Metrics} 
We used the SWE-bench metrics which is the percentage of task instances that are correctly solved by the model. In order to judge whether or not the model correctly solves an individual task we will use the following scoring method:

\begin{table}[ht]
\centering
\caption{SWE-bench Evaluation}

\begin{tabular}{lc}
\toprule
\textbf{Result} & \textbf{Score} \\
\midrule
Fails at any step & 0 \\
Correctly completes all steps & 1 \\
\bottomrule
\end{tabular}
\label{tab:swe-bench_eval}
\end{table}

If the evaluation produces incorrect outputs for any step in the task then the entire task is treated as a failure. In order to be counted as a success, the evaluation must succeed at every step of and pass all the associated test cases. Taking the total number of successful tasks over the number of attempted tasks will serve as our primary metric.

\begin{figure*}[h]
    \centering
    \includegraphics[width=\textwidth]{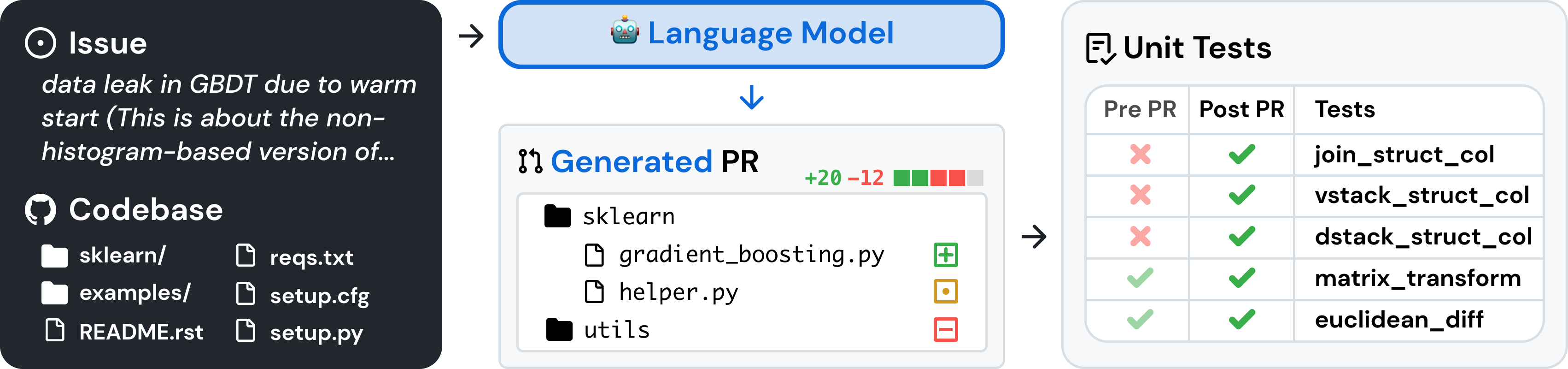}
    \caption{SWE-bench Evaluation \cite{jimenez2024swebench}}
    \label{fig:swe_bench}
\end{figure*}

\subsubsection{Evaluation} 
We will execute generated patches against the corresponding task instances of the benchmark to determine whether or not it resolves the associated Github issue. The SWE-bench refers to such patch generations as the \textit{prediction patch}.
The benchmark framework performs the following steps for testing:

\begin{enumerate}
    \item Installs repository at base commit according to task instructions
    \item Applies test patch, \textit{prediction patch} and run tests
    \item Checks prediction logs to see the pass/fail status of each test
\end{enumerate}

\subsection{Fine-tuning}
We loaded the CodeLlama model with \textit{FastLanguageModel} loader from  the unsloth\cite{han2023unsloth} library that extends Hugging Face's Parameter-Efficient Fine-Tuning (PEFT) \cite{peft} which provides several performance optimizations for training and inference. 

\begin{table}[hb]
\centering
\caption{CodeLlama training parameters.}
\label{tab:parameters}
\begin{tabular}{lc}
\toprule
\textbf{Parameter} & \textbf{Value} \\
\midrule
Quantization & 4-bit \\
Learning rate & 2e-4 \\
Optimizer & AdamW 8-bit \\
Warmup ratio & 0.05 \\
Number of epochs & 1 \\
Max seq length & 16384 \\
Weight decay & 0.01 \\
Per device train batch size  & 4 \\
Grad accumulation Steps & 4 \\
\bottomrule
\end{tabular}
\end{table}

The trainer was setup with the parameters shown in table \ref{tab:parameters}. Additionally, we applied the following techniques:

\begin{enumerate}
    \item Supervised training: Supervised Fine-tuning Trainer \textit{SFTTrainer} from Hugging Face's Transformer Reinforcement Learning (TRL) \cite{vonwerra2022trl} library.
    \item Quantization: We used a pre-quantized 4-bit model to reduce memory usage.
    \item Lower Ranking Adaptation(LoRA): Using QLoRA\cite{dettmers2023qlora} we only updated 1 to 10\% of all parameters.
    \item Rotary Positional Embedding(RoPE) Scaling: the of RoPE \cite{su2022roformer} scaling using Kiao Ken's method\cite{ken2023} made the context window flexible.
\end{enumerate}

\subsection{Inference via API}
To mitigate the computational challenges and time constraints associated with the experiments the Groq API was utilized for both \textbf{Mixtral-8x7B} and \textbf{Llama2 70B}, achieving an impressive average throughput of 500 tokens per second; A large boost in the generation speed of patches. Looking forward, we are encouraged by the expected arrival of Language Processing Units (LPU) \cite{Abts2022TPU} Inference Engines which promise to significantly advance the field by facilitating the adoption of frameworks like ToT across a broad spectrum of applications.

\section{Results} 

\begin{table}[htb]
\centering
\caption{SWE-bench Lite results.}
\label{tab:results}
\begin{minipage}{0.5\textwidth}
\centering
\begin{tabular}{lcc}
\toprule
\textbf{Model} & \multicolumn{2}{c}{\textbf{Success Rate (\%)}} \\
\cmidrule(r){2-3}
 & \textbf{IO} & \textbf{ToT}\\ 
\midrule
CodeLlama 34B\textsuperscript{+} & 0 & 0 \\
Llama-2 70B\textsuperscript{+} & 0 & 0\\
Mixtral-8x7B\textsuperscript{+} & 0 & 0 \\
\midrule
Llama-3 70B Instruct\textsuperscript{*} & 0 & 0 \\
\bottomrule
\end{tabular}
\vspace{0.5\baselineskip} 
\end{minipage}
\footnotesize 
\textsuperscript{+}33\% and
\textsuperscript{*}50\% of dataset respectively.
\end{table}

We conducted the experiments using a subset of 100 instances, representing 33\% of the full dataset from the SWE-bench\_Lite benchmark. To test our last model \textbf{Llama-3 70B Instruct}, we ran a bigger subset of 150 instances representing 50\% of the dataset. These sample sizes were selected to provide a representative snapshot of ToT's performance, while also considering the computational constraints and resource requirements associated with evaluating the models on the complete dataset.

The three models performed as poorly as input-output (IO) prompting, this is shown in table \ref{tab:results}. We observed that while all the generated git patches were syntactically correct (demonstrating the in-context learning capabilities of the LMs), none of them were able to successfully pass the benchmark.

\begin{table}[htb]
\centering
\caption{Accepted patches}
\label{tab:accepted_patches}
\begin{minipage}{0.5\textwidth}
\centering
\begin{tabular}{lcc}
\toprule
\textbf{Model} & \multicolumn{2}{c}{\textbf{Accepted Patches (\%)}} \\
\cmidrule(r){2-3}
 & \textbf{IO} & \textbf{ToT}\\ 
\midrule
CodeLlama 34B\textsuperscript{+} & 0 & 0 \\
Llama-2 70B\textsuperscript{+} & 0 & 0\\
Mixtral-8x7B\textsuperscript{+} & 0 & 0 \\
\midrule
Llama-3 70B Instruct\textsuperscript{*} & 1 & 10 \\
\bottomrule
\end{tabular}
\vspace{0.5\baselineskip} 
\end{minipage}
\footnotesize 
\textsuperscript{+}33\% and
\textsuperscript{*}50\% of dataset respectively.
\end{table}

As presented in Table \ref{tab:accepted_patches}, it is noteworthy that for \textbf{Llama-3 70B Instruct}, 10\% of the generated patches were \textit{accepted}. However, these patches subsequently failed to satisfy certain test cases within the test suite. In other words, these patches subsequently failed to meet the benchmark success criteria. In contrast, with \textbf{CodeLlama 34B}, \textbf{Llama-2 70B}, and \textbf{Mixtral-8x7B} all the proposed patches were \textit{rejected} outright, precluding any chance for testing.

\section{Analysis} 
The findings indicate that Tree of Thoughts (ToT) was not effective for the specific task that was examined. 
This research acknowledges several limitations within the experimental setup that may have impacted the efficacy of the ToT framework. Specifically, we identify the following weaknesses:

\begin{enumerate}
    \item The use of a relatively shallow thought process tree, consisting of only two thought steps. This was problematic because it did not allow the complex tasks asked of the model to be decomposed into smaller, more manageable sub-tasks that individual reasoning steps could be applied to. 
    \item Providing only the repository name, problem statement, and git commit is insufficient for the model to fully comprehend and address the requirements of the task. However, as shown in table \ref{tab:swe_lite_leaderboard} many complex tasks cannot be accomplished through a single step or a solitary tool invocation. Even with RAG conducting similarity searches over the contents of the sizable Git repository proved to be an ineffective way of supplying the model with the necessary task-specific information. Allowing the model access to explore and examine the file contents of the associated GitHub repository could have facilitated a more informed and accurate approach.
    \item The validation of the individual thought steps was conducted through a voting mechanism, rather than comparing the outputs to a defined \textit{ground truth}. Incorporating a symbolic validation component, similar to the utilized by SWE-bench itself, could have provided a more robust means of evaluating the correctness of the generated patches.
\end{enumerate}

\section{Conclusion}

Despite the limitations observed in the current experimental setup, we hypothesize that with a redesigned approach and addressing the identified weaknesses, the potential of the Tree of Thoughts (ToT) framework could be better realized and more effectively demonstrated. Specifically, a setup with \textit{agentic design patterns}\cite{Ng2024} where the large language model is:
\begin{enumerate}
    \item Leveraged to autonomously break down the objective into smaller sub-tasks. This allows the model to dynamically determine the optimal sequence of steps required to accomplish the resolution of the GitHub issue. 
    \item Provided with access to tools via \textit{function calling}\cite{kim2024llm}. This enables the model to independently make requests for the purpose of gathering information, taking action, or manipulating data. For example, being able to do a code \texttt{search} in the git repository and \texttt{open} sections of a file like the Agent-Computer Interface (ACI) introduced by SWE-agent.
    \item Provided with code search that is project structure aware like AutoCodeRover\cite{zhang2024autocoderover}. Instead of searching over files by plain string matching, the model can search for relevant code context (functions/classes) in the syntax tree.
\end{enumerate}
Additional improvements are related to the patch generation, such as:
\begin{enumerate}
    \item Freeing the large language model from the direct responsibility of generating the code patches. Instead, leveraging the use of tools via function calling to handle the patch generation, allowing the model to focus on planning, debugging and code generation.
    \item Incorporating a reliable ground truth for validating the generated patches.
\end{enumerate}
Finally, since the model no longer generates unified patches, fine-tuning as an optimization strategy becomes less effective.

\section*{Known Project Limitations}

\textbf{Speed and Cost}: It is important to note that while ToT may enhance decision-making and problem-solving capabilities of large language models, this sophistication can result in slower processing times due to the additional computational steps involved. The evaluation of multiple reasoning paths and self-assesses choices, inherently demands more resources and compute time, including a notable rise in prompt and generation tokens. However, the trade-off for this slower speed and higher cost is a potential increase in the accuracy and relevance of the outcomes, particularly in complex tasks. Adjustments to the framework's parameters can offer some mitigation of these issues, allowing users to make a balance between speed/cost and accuracy.\\
\textbf{Search Methods}: This research leverages the use of classical search algorithms, such as Breadth-First Search (BFS) and the current setup can be considered a form of heuristic search, akin to the A* algorithm, where the heuristic at each search node is provided by the large language model's own self-assessment of the generated thought.
While this search strategy is straightforward to implement, it represents a relatively naive approach. We anticipate that the use of more advanced search algorithms, such as Monte Carlo Tree Search (MCTS), could potentially yield improved results. This expectation is informed by prior work, such as the research conducted by \cite{hao2023reasoning} with Reasoning viA Planning.

\section*{Authorship Statement}
Ricardo La Rosa conceptualized the core experiments and hypothesis, conducted the majority of the experiments, analyzed the data, and wrote the core of the manuscript. Corey Hulse helped conduct validation of the experiments and oversaw final edits of deliverables. Bangdi Liu provided critical feedback, and assisted with revisions to the manuscript and previous deliverables. All authors read and approved the final version of the manuscript.

\section*{Acknowledgements}
We express our sincere gratitude to Christopher Potts, Petra Parikova, and our course facilitator, Jonathan Gomes Selman, for their valuable contributions and support throughout the duration of Stanford's XCS224U course. We are grateful for their generous availability and support.

\bibliography{anthology,custom}

\appendix
\section{Prompts}\label{sec:appendix}
\begin{lstlisting}[caption={Plan prompt}]
plan_prompt = '''Given the Repository url, Base commit and Problem statement of a github issue. Please write a plan to solve it.
Your output must be of the following format:

Plan:
Your plan here.

{input}
''' 
\end{lstlisting}
\begin{lstlisting}[caption={Patch prompt}]
patch_prompt = '''Given the Repository url, Base commit, Problem statement of a github issue and a plan. Please write a correct git patch to solve it.

Your output must be of the following format:

Patch:
```diff
Your patch here.
```

The patch file should be in the unified diff format. Example:

```diff
diff --git a/file.py b/file.py
--- a/file.py
+++ b/file.py
@@ -1,27 +1,35 @@
 def euclidean(a, b):
-    while b:
-        a, b = b, a % b
-    return a
+    if b == 0:
+        return a
+    return euclidean(b, a % b)
```

{input}
''' 
\end{lstlisting}

\begin{lstlisting}[caption={Vote prompt}]
vote_prompt = '''Given an instruction and several choices, decide which choice is most promising. Analyze each choice in detail, then conclude in the last line "The best choice is {s}", where {s} the integer id of the choice.'''
\end{lstlisting}
\begin{lstlisting}[caption={Score prompt}]
score_prompt = '''Analyze the following patch, then at the last line conclude "Therefore the correctness score is {s}", where {s} is an integer from 1 to 10.'''
\end{lstlisting}
\end{document}